\documentclass[11pt,twoside]{article}
\usepackage{./asp2014}

\aspSuppressVolSlug
\resetcounters

\begin{document}

\title{Science with an ngVLA: \\ Dust growth and dust trapping in protoplanetary disks}

\author{Nienke van der Marel, Brenda Matthews \vspace{1mm}
\affil{Herzberg Astronomy \& Astrophysics Programs, National Research Council of Canada, Victoria, BC, Canada}} %\email{astro@nienkevandermarel.com}}

\author{Ruobing Dong \vspace{1mm}
\affil{ Department of Physics \& Astronomy, University of Victoria, Victoria, BC, Canada}} %\email{AuthorEmail@email.edu}}

\author{Tilman Birnstiel \vspace{1mm}
\affil{University Observatory, Faculty of Physics, Ludwig-Maximilians-Universitat Munchen, Munich, Germany;}}

\author{Andrea Isella \vspace{1mm}
\affil{Department of Physics \& Astronomy, Rice University, Houston, TX, USA;}}
 %\email{AuthorEmail@email.edu}}}

% This section is for ADS Processing.  There must be one line per author.
\paperauthor{Nienke van der Marel}{astro@nienkevandermarel.com}{ 0000-0003-2458-9756}{National Research Council of Canada}{NRC Herzberg Astronomy \& Astrophysics Programs}{Victoria}{BC}{V9A 2E7}{Canada}
\paperauthor{Brenda Matthews}{bcmatthews.herzberg@gmail.com}{}{National Research Council of Canada}{Herzberg Astronomy \& Astrophysics Programs}{Victoria}{BC}{V9A 2E7}{Canada}
\paperauthor{Ruobing Dong}{rbdong@gmail.com}{}{University of Victoria}{Department of Physics \& Astronomy}{Victoria}{BC}{}{Canada}
\paperauthor{Til Birnstiel}{til.birnstiel@lmu.de}{}{LMU}{}{Munich}{}{}{Germany}
\paperauthor{Andrea Isella}{isella@rice.edu}{}{Rice University}{Department of Physics \& Astronomy}{Houston}{TX}{}{USA}

%Please include a brief abstract that will be used by ADS for searching purposes.  
\begin{abstract}
ALMA has revolutionized our view of protoplanetary disks, revealing structures such as gaps, rings and asymmetries that indicate dust trapping as an important mechanism in the planet formation process. However, the high resolution images have also shown that the optically thin assumption for millimeter continuum emission may not be valid and the low values of the spectral index may be related to optical depth rather than dust growth. Longer wavelength observations are essential to properly disentangle these effects. The high sensitivity and spatial resolution of the next-generation Very Large Array (ngVLA) will open up the possibilities to spatially resolve disk continuum emission at centimeter wavelengths and beyond, which allows the study of dust growth in disks in the optically thin regime and further constrain models of planet formation.
\end{abstract}

\section{Introduction}
The planet formation process remains one of the major puzzles in modern-day astronomy. Planets are known to form in protoplanetary disks of gas and dust around young stars \citep{WilliamsCieza2011}. A large amount of research has been conducted in the second half of the twentieth century, but many questions remain unanswered. In particular, planet formation is hindered by a number of growth barriers, according to dust evolution theory, while observational evidence indicates that somehow these barriers must have been overcome. The core accretion process, where the accretion of dust particles and planetesimals results in solid cores of $\sim$Earth mass, followed by runaway gas accretion, is a promising mechanism to form the range of planets that are seen in both our Solar System and beyond \citep[e.g.][]{Winn2015}. However, the core accretion process requires the growth of planetesimals from submicron-sized dust grains that are seen in the interstellar medium (ISM). These first steps in dust growth in protoplanetary disks, where the dust evolution is governed by the drag forces between dust and gas, have proven to be one of the most challenging parts in the planet formation process \citep[e.g.][and references therein]{Testi2014}. 

%Planets are thought to form in protoplanetary disks of gas and dust around young stars, where the disk itself is a by-product of the star formation process due to conservation of angular momentum. Giant planets of $\sim$ Jupiter mass must form before the gas in the disk is dissipated, which is thought to happen within the first 5-10 million years of the life of the disk \citep{WilliamsCieza2011}. Although gravitational instability or fragmentation allow a rapid concentration and formation of Jupiter like planets \citep[e.g.][and references therein]{Helled2014}, protoplanetary disk masses are generally too low to make this mechanism a common pathway for the formation of planetary systems. This is a problem in particular considering the observed diversity in exoplanetary masses \citep[e.g.][]{Winn2015}, which cannot all be produced by fragmentation. The core accretion process, where the accretion of dust particles and planetesimals results into solid cores of $\sim$Earth mass, followed by runaway gas accretion, is a more promising mechanism to form the range of planets that are seen in both our Solar System and beyond. However, the core accretion process requires the growth of planetesimals from submicron-sized dust grains that are seen in the interstellar medium (ISM). These first steps in dust growth in protoplanetary disks, where the dust evolution is governed by the drag forces between dust and gas, have proven to be one of the most challenging parts in the planet formation process \citep[e.g.][and references therein]{Testi2014}. 

In the last two decades, clear evidence has been found for dust grain evolution in disks beyond the grain sizes in the interstellar medium, through %analysis of mid-infrared spectra of silicate features \citep[e.g.][]{vanBoekel2003,Kessler2006} and 
(interferometric) millimeter observations of disks \citep[e.g.]{Beckwith1991,Testi2003,AndrewsWilliams2005,Natta2007, Isella2009,Ricci2010a}, indicating the presence of millimeter and even centimeter-sized particles in the outer regions of protoplanetary disks through multi-wavelength continuum observations and measurements of the spectral index $\alpha_{mm}$ (millimeter flux $F_{\nu}\sim\nu^{\alpha_{mm}}$). The value of $\alpha_{mm}$ can provide information on the particle size in protoplanetary disks \citep[see][and references therein]{Testi2014}. For (sub)micrometer-sized dust, such as found in the ISM (interstellar medium), $\alpha_{mm}$ is typically 3.5--4.0, but when dust grows to millimeter sizes, $\alpha_{mm}$ is expected to decrease to 2--3 \citep{Draine2006,Ricci2010a}. When the dust emission is optically thin and in the Rayleigh-Jeans regime, the observable $\alpha_{mm}$ can be related to the dust opacity index $\beta = \alpha-2$, with $\beta<1$ for millimeter grains and larger \citep{Natta2004}. Furthermore, $\beta$ is  related to the dust opacity of a certain dust grain size population $n(a)$ as $\kappa_{\nu} \propto \nu^{\beta}$, where $\beta$ is mostly sensitive to the maximum grain size $a_{\rm max}$ \citep{Draine2006}. 
%and related to the optical depth as $\tau_{\nu}\propto\nu^{\beta}$. 

\begin{figure}[!ht]
\includegraphics[width=\textwidth]{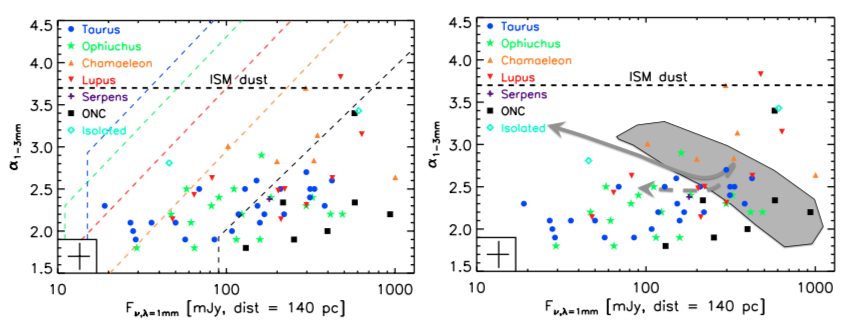}
\caption{Spectral index $\alpha_{mm}$ between 1.1 and 3 mm against 1.1 mm flux for disks in various nearby star forming regions. The dashed diagonal lines (left panel) mark the sensitivity limits of the various surveys. The grey area in the right panel illustrates the predictions of dust evolution models \citep{Birnstiel2010}, and the arrows indicate the effect of dust evolution including radial drift (solid) and pressure traps (dashed) \citep{Pinilla2012a}. Figure taken from \citet{Testi2014}.}
\label{fig:testifigure}
\end{figure}

Initially, grain growth in protoplanetary disks was primarily based on unresolved or marginally resolved millimeter data in surveys of star forming regions, showing $\alpha_{mm}$ values well below the ISM value of 3.5--4.0 (see Figure \ref{fig:testifigure}), indicating millimeter-sized particles in the outer disk. For a long time, it remained unclear how these particles could be present, due to the so-called \emph{radial drift} problem,
%The main issue in explaining the presence of these dust particles is the so-called \emph{radial drift} problem, 
which results in a rapid inward drift of dust particles in a disk when grown to millimeter sizes \citep{Whipple1972,Weidenschilling1977}. For a disk with a smooth radial density profile, both the gas surface density and temperature, and thus the pressure, decrease radially outward. This additional pressure support results in a slightly sub-Keplerian orbital velocity for the gas. 
%In contrast, the dust particles are not pressure supported and should thus move in Keplerian orbits. The small velocity difference between gas and dust thus causes the dust particle to feel a head wind or friction, resulting in a very efficient deceleration of the dust particle, consequently loosing angular momentum and effectively spiraling inwards. 
In contrast, dust particles are not pressure supported and want to orbit at Keplerian speed, but as they are embedded in a sub-Keplerian gas disk, they are forced to orbit at lower speed which leads to a removal of angular momentum from the particles to the gas, causing the particles to spiral inward. The dust particles are thus experiencing a drag force by the gas, depending on its Stokes number St, which describes the coupling of the particles to the gas and depends on the dust particle size and the local gas surface density \citep[see e.g.][]{Brauer2008,Birnstiel2010}. For small dust particles, St$\ll$1, they are strongly coupled to the gas and do not drift inwards, but radial drift becomes significant when the particle size increases and reaches its strongest value when St = 1. In practice, millimeter-dust particles in the outer disk can have St values close to unity and are expected to drift inwards on time scales as short as 100 years, hence further dust growth is hindered. An additional problem is \emph{fragmentation}: while low-velocity collisions lead to particle growth, high-velocity impacts lead to destruction according to experimental and theoretical work on dust interaction \citep{Blum2000}. The combination of the radial drift and fragmentation problems is also called the `meter-size barrier' because a one-meter size object at 1 AU drifts efficiently inwards limiting further growth, and equivalently dust particles in the outer disk cannot grow beyond millimeter sizes. 

A proposed solution to explain the presence of larger dust grains in disks as the start of planet formation, is a so-called \emph{dust trap}, where dust particles are being `trapped' in local pressure maxima in the outer disk \citep{Whipple1972,KlahrHenning1997,Rice2006,Brauer2008,Pinilla2012a}. Such a pressure maximum or pressure bump can arise as a radial dust trap at the edges of dead zones \citep[e.g.][]{Varniere2006}, at the edges of gas gaps cleared by planets \citep{Zhu2012,Pinilla2012b}, in zonal flows \citep{Johansen2009, Pinilla2012a} or as azimuthal dust traps in long-lived vortices \citep[e.g.][]{BargeSommeria1995, KlahrHenning1997}, which can be the result of a Rossby Wave Instability of a radial pressure bump \citep[e.g.][]{Lovelace1999,WolfKlahr2002, Lyra2009, Regaly2012}. Pressure maxima as a solution for dust growth beyond millimeter sizes were proposed theoretically, but without spatially resolved information on the dust grain size distribution within the disk this phenomenon remained speculative.

The  \emph{Atacama Large Millimeter/submillimeter Array} (ALMA) in Chile, which started operations in 2012, has revolutionized our view of protoplanetary disks, in particular on the existence of dust traps, starting with the discovery of the highly asymmetric millimeter-dust concentration in Oph IRS 48 \citep{vanderMarel2013}. Whereas the gas and micrometer-dust are distributed along a ring the millimeter emission showed evidence for dust trapping in a vortex, thought to be generated by a Rossby Wave Instability of a radial pressure bump. Follow-up observations at centimeter wavelengths with the VLA confirmed that the azimuthal width of the dust trap decreased with wavelength \citep{vanderMarel2015vla}, as predicted by dust trapping models \citep[e.g.][]{Birnstiel2013}. Oph IRS 48 is a so-called transitional disk, a disk with a cleared inner dust cavity \citep[e.g.][and references therein]{Espaillat2014}, which were of particular interest in dust evolution predictions for keeping millimeter-sized dust grains in the outer disk \citep{Pinilla2012b}. Other transition disks imaged with ALMA show a range of azimuthally symmetric and asymmetric dust rings \citep{Casassus2013,Zhang2014,Perez2014,vanderMarel2015-12co,Pinilla2017-sr24s}, confirming their ring-like morphologies as seen with previous interferometers such as CARMA and the SMA \citep[e.g.][]{Brown2009,Isella2010mwc758,Andrews2011}. The existence of radial dust traps was primarily indicated by the difference in distribution of the gas (as traced by CO isotopologues) and millimeter dust \citep[e.g.][]{Bruderer2014,SPerez2015,vanderMarel2016-isot,Dong2017,Fedele2017} and the difference with the micrometer-sized dust grains \citep{Garufi2013,Pinilla2015}. Multi-wavelength ALMA observations between 1.3 and 0.45 millimeter provided a hint of radial trapping in the SR~21 and SR~24S disks through spatially resolved $\alpha(r)$ which decreased to $<$3 in the ring emission \citep{Pinilla2014beta,Pinilla2017-sr24s}, and azimuthal changes in $\alpha_{mm}$ were seen in the asymmetries in HD~142527 \citep{Casassus2015} and HD~135344B \citep{Cazzoletti2018hd13}. Combining ALMA with VLA data at centimeter wavelengths results in further evidence for dust trapping in MWC~758 \citep[e.g.][]{Marino2015}, but the low spatial resolution and sensitivity limits the calculation of a spatially resolved $\alpha_{mm}(r)$ map. 

Another spectacular ALMA result in this context is the imaging of disks at ultra high angular resolution of $\sim$20 mas or a few AU at the distance of nearby star forming regions. The mind-blowing image of HL~Tau \citep{HLTau2015} of the ALMA Long Baseline Campaign has revealed that even dust disks without inner cavity may exist of ring-like structures, which a remarkable similarity to the predictions of \citet{Pinilla2012a} of local pressure maxima that are required for the explanation of millimeter-sized dust in the outer disk. The HL~Tau image was quickly followed by other multi-ring disks, such as TW~Hya \citep{Andrews2016} and HD~163296 \citep{Isella2016}. %Also several transition disks turned out to consist of multiple rings, e.g. HD~97048 \citep{vanderPlas2017}, HD~169142 \citep{Fedele2017} or even a combination of rings and asymmetries in HD~135344B \citep{vanderMarel2016-spirals}, V1247 Ori \citep{Kraus2017} and MWC~758 \citep{Dong2018}. An ALMA Large Program is on the way to reveal even more rings and other substructures in a large number of the brightest primordial disks (Andrews et al. in prep.). 
Explanations for the dust rings range from planets carving gaps \citep{LinPapaloizou1979}, snow lines \citep{Zhang2015}, sintering \citep{Okuzumi2016} and secular gravitational instabilities \citep{Takahashi2016}. If the dust rings are indeed caused by planets, trapping is expected to occur. Multi-wavelength observations indeed reveal radial variations of $\alpha_{mm}$ along the gaps and rings in HL~Tau and TW~Hya \citep{Carrasco2016,Tsukagoshi2016}, but the evidence is still marginal with the currently available data. %Snow lines could explain some of the ring locations without the need for trapping, but this suggests some correlation between the ring locations and the stellar properties which is not observed (van der Marel et al. in prep., Huang et al. in prep.). Even more for these narrow rings, optical depth is an issue in the interpretation of the dust emission. 
On the other hand, evidence for radial drift is evident in observations. SMA observations already revealed evidence for a segregation between the dust and gas in the IM Lup disk \citep{Panic2009}, with the gas outer radius being twice as large as that of the dust. Multi-wavelength continuum observations show a decrease of particle grain size with radius for a number of primordial disks \citep{Perez2015,Tazzari2016,Tripathi2018}. Clearly, in lack of one or more dust traps in the outer disk, dust particles do drift inwards to the nearest pressure maximum. On the other hand, this implies that any extended dust disk must in fact consist of one or more pressure bumps, to keep the dust particles away from the center of the disk.

\section{The optical depth problem}
The results described above appear very promising in explaining the presence of large dust grains in protoplanetary disks as the start of the planet formation process, but they suffer from a major problem: the spectral index $\alpha_{mm}$ can only be related to $\beta$ in a simple linear fashion for optically thin emission. The optical depth $\tau_{\nu}(r)$ is defined as 
\begin{equation}
\tau_{\nu}(r) = \int_{-\infty}^\infty \rho \kappa_{\nu} ds = \frac{\kappa_{\nu} \Sigma(r)}{\cos i}
\end{equation}
with $\Sigma(r)$ the dust surface density at radius $r$ and $i$ the inclination and the emitted flux $F_{\nu}$ integrated over a disk area can be computed as:
\begin{equation}
F_{\nu} = \frac{\cos i}{d^2} \int {B_{\nu} (T(r)) \Big(1-e^{-\tau_{\nu}(r)}\Big) 2\pi rdr}
\end{equation}
with distance $d$, Planck function $B$ at temperature $T$, which reduces to $F_{\nu} \propto B_{\nu}(T) \propto \nu^2$ for optically thick emission and  $F_{\nu} \propto \kappa_{\nu}B_{\nu}(T) \propto \nu^{2+\beta}$ for optically thin emission ($\tau_{\nu} \ll 1$). This immediately illustrates that the spectral index $\alpha_{mm}$ is 2 (and thus $\beta\sim 0$) for optically thick emission regardless of the grain size, and a low value of $\beta$ does not automatically imply large dust grains. The optical depth is generally estimated by comparing the measured brightness temperature $T_b$ ($T_b\sim I_{\nu} = \frac{F_{\nu}}{d\Omega}$) with the solid angle $d\Omega$ and the physical dust temperature $T_{\rm dust}$ as calculated from a radiative transfer model. Typical disk observations of millimeter interferometry do show lower values for $T_b$ than $T_{\rm dust}$, indicating $\tau_{\nu}\sim 0.5$ or lower, but if the optically thick emission is concentrated on size scales smaller than the resolution, optical depth effects cannot be ruled out as an explanation for the low $\beta$ values based on millimeter emission \citep{Ricci2012, Tripathi2017}. This is generally measured using the filling factor $f$, the fraction of the disk area which is optically thick. Thus, the evidence for dust growth and dust trapping based on spectral index values from millimeter observations may not be reliable. Furthermore, if the assumption of optically thin dust emission is no longer valid, this also has potentially large consequences for disk dust mass estimates, which are generally taken as a linear relation between $M_{\rm dust}$ and integrated millimeter flux $F_{\nu}$ as:
\begin{equation}
M_{\rm dust} = \frac{F_{\nu}d^2}{\kappa_{\nu}B_{\nu}(T)}
\end{equation}
If the dust emission is optically thick, the derived disk dust mass is only a lower limit. Even in high resolution ALMA observations, the millimeter disk emission often remains unresolved (e.g. in $>$30\% of the disks in the Lupus survey which has a spatial resolution of 0.25" or 20 AU radius, \citet{Ansdell2018}), implying that the majority of disks are much smaller than previously thought, and their emission is potentially optically thick. However, even for partially resolved extended disks the integrated flux may be dominated by optically thick emission from either rings or from the inner part of the disk where the dust concentrates as a result of radial drift in absence of pressure maxima in the outer disk \citep{Tripathi2018}. The high spatial resolution of ALMA is a crucial factor here as high resolution observations have demonstrated that dust emission is concentrated in much narrower rings than previously thought. 
%: rings in transition disks are much narrower than previously thought, and primordial disks are not large smooth solid areas, but consist of many narrow dust rings, which are each likely optically thick as the emission originates from a much smaller surface. 

\section{The role of the next-generation VLA}
In order to study dust growth in protoplanetary disks, spatially resolved multi-wavelength observations in the optically thin regime are crucial. The next-generation Very Large Array (ngVLA) offers high angular resolution ($\sim$10 mas) at long wavelengths (17-93 GHz or 3-18 mm) where $\tau_{\nu} \ll 1$. The sensitivity ($\sim0.3 \mu$Jy/beam for 1 hour integration at 40 GHz) allows quality observations at high signal-to-noise, resulting in clear images of the structures in disks. The current VLA does not have the resolution or sensitivity to achieve this result in a reasonable amount of time. Of particular importance for dust growth studies is the spatially resolved spectral index $\beta_{cm}$ which can be calculated across optically thin wavelengths. This opens up a large range of possibilities to answer questions on the origins and mechanisms of the structures in disks observed by ALMA. 

\subsection{Azimuthal asymmetries}
The azimuthal asymmetries, though to be dust trapping vortices, have gained a large interest from the community, but a lot of questions regarding their origin remain unclear. A spatially resolved $\beta_{cm}(r,\phi)$ can help to disentangle azimuthal trapping from other proposed effects to explain dust asymmetries such as eccentricities \citep{Ataiee2014,Ragusa2017}, as eccentric disks would not show an azimuthal dependence of $\beta_{cm}$. Furthermore, the measurement of $\beta_{cm}(r,\phi)$ can constrain the trapping efficiency in the disk, and in combination with estimates of the gas surface density, lead to independent constraints on the viscosity and turbulence \citep{Birnstiel2013,LyraLin2013}, which tends to diffuse the dust in the vortex. Quantitative estimates of the turbulence can further be used to estimate the time scales of dust trapping in a vortex and relate them to their observability. For AB Aur, \citet{Fuente2017} claim a decaying vortex due to the azimuthal extended emission at longer wavelengths (in contrast to dust trapping predictions), but optical depth effects need to be excluded to confirm this claim.

Another interesting consequence of trapping in a vortex is a size segregation due to the vortex's self-gravity: whereas smaller grains will be trapped in the center of the vortex, larger grains are expected to be trapped \emph{ahead} in the azimuthal direction \citep{BaruteauZhu2016}: a shift was indeed observed in the asymmetry in HD~135344B in ALMA observations, but in the opposite direction \citep{Cazzoletti2018hd13}. In order to rule out any effects due to optically thick emission, ngVLA observations at similar angular resolution, can test the predictions of \citet{BaruteauZhu2016} in a large number of disks. %On the other hand, the asymmetry in AB Aur was shown to be \emph{more} extended at longer wavelengths \citep{Fuente2017}, opposite to the dust trapping predictions. This was interpreted as a time scale effect: as the vortex decays, dust particles diffuse on different time scales out of azimuthal trapping, depending on their particle size. 

\begin{figure}[!ht]
\includegraphics[width=\textwidth,trim=20 20 20 20]{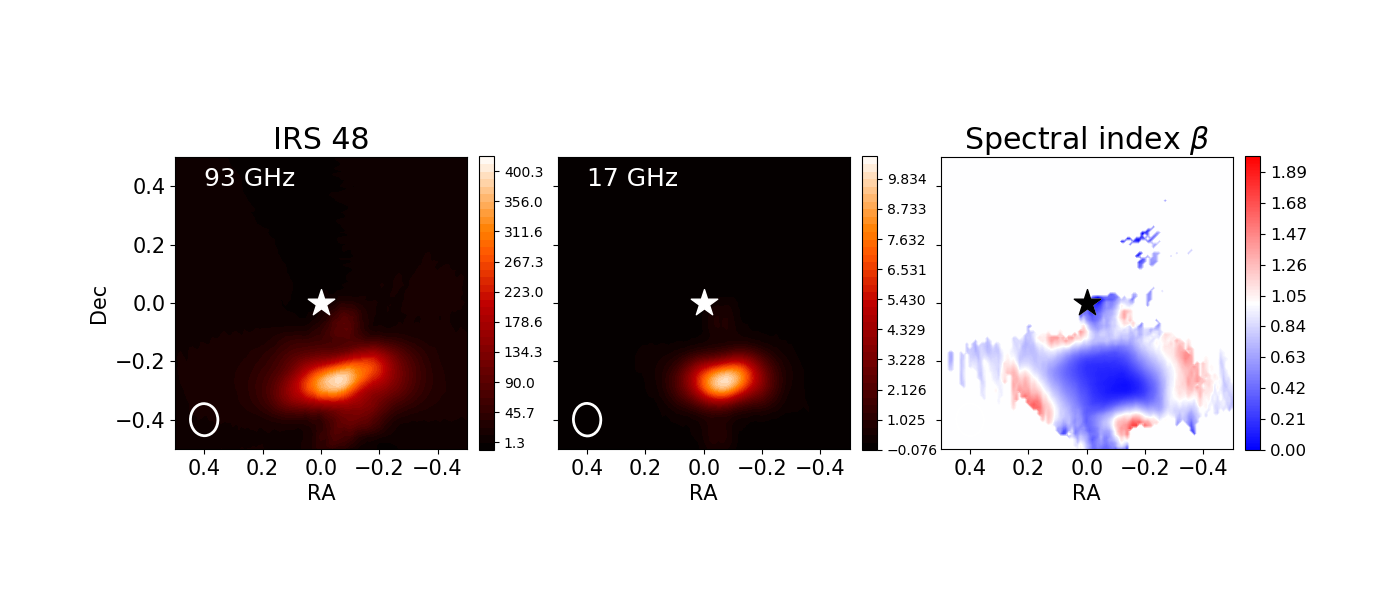}
\caption{Simulated images of the spectral index $\beta(r,\phi)$ for the azimuthal aymmetry in Oph IRS 48 at centimeter wavelengths, as expected from ngVLA observations at 97 and 17 GHz in an 2 hour integration. Flux units are given in $\mu$Jy/beam. Oph IRS 48 is expected to show a decrease of $\beta$ in the center due to dust growth. }
\label{fig:simulation}
\end{figure}

As a demonstration, we have created intensity profiles of a highly asymmetric disks (Oph IRS 48) based on its observed morphology with ALMA and VLA and scaled to their expected centimeter flux at 97 and 17 GHz. In Figure \ref{fig:simulation} we demonstrate that the spatially resolved $\beta_{cm}$ as calculated from 3 and 18 millimeter emission can be measured with an accuracy of $\sim$0.1 in high SNR observations with the ngVLA in 2 hours per setting, using the CASA simulator and the ngVLA configuration. 

\subsection{Multi-ring disks}
Spatially resolving the multi-ring systems such as HL~Tau in multiple centimeter wavelengths will help to constrain their origin, which is currently completely unclear: cleared gaps by planets is a tempting explanation, but this requires (giant) planets at tens or even hundreds of AU away from the star, which are very rare in exoplanet demographics \citep{Winn2015,Bowler2016} and difficult to explain in reasonable time scales in planet formation theory \citep{Helled2014}. The edges of planet gaps are pressure maxima where dust is expected to be trapped; a measure of $\beta_{cm}(r)$ can immediately tell if the rings are consistent with dust traps. If not, other explanations for the origin of the rings will need to be considered.

\subsection{Radial drift}
In disks without pressure maxima in the outer disk, the larger dust particles are expected to drift inwards and form a sharp outer continuum edge \citep{Birnstiel2014}. Although this effect has been demonstrated through comparison between the millimeter dust emission and the much more extended distribution of the gas as traced by the CO \citep[e.g.][]{Panic2009,Andrews2012,deGregorio2013,Ansdell2018}, a quantification of the drift and associated time scales remain unexplored due to the uncertainties introduced by the high optical depth in the inner part of the dust disk \citep[e.g.][]{Perez2015,Tazzari2016,Tripathi2018}. Measuring $\beta_{cm}(r)$ in optically thin wavelengths will set better constraints on the grain size distribution within the disk and will help to constrain dust evolution models in providing better estimates for e.g. drift time scales and fragmentation.

\subsection{Disk masses}
Disk dust masses are generally derived from disk-integrated millimeter fluxes using the assumption of optically thin emission, but it is quite likely that this assumption is not valid, implying that many of the derived disk masses (usually calculated with the dust mass and a gas-to-dust ratio of 100) are lower limits. This has important consequences for our understanding of the dynamics in disks: gravitational instability \citep{Boss1997} may occur for disk masses $>$10\% of the stellar mass, which is generally well above our current estimates of the disk mass. However, if large amounts of the millimeter emission is optically thick, these disk masses are severely underestimated and gravitational instabilities could be important for disk fragmentation, dynamics and giant planet formation than currently thought. Large disk surveys with the ngVLA of nearby star forming regions can provide us with flux measurements of hundreds of disks, similar to current ALMA surveys. The sensitivity of the Lupus ALMA disk survey \citep[$\sim0.3 M_{\rm Earth}$ as 3$\sigma$ detection limit][]{Ansdell2018} can be reached in 2 minutes per source with the ngVLA at 40 GHz, allowing full snapshot surveys in only a few hours.

% examples: 
% - azimuthal trapping: trapping efficiency: viscosity, peak shift, diffusion vortex, eccentricity vs trapping
% - multi-rings: radial trapping?
% - drift: inner part of the disk
% - general: disk mass

\section{Conclusions}
Whereas ALMA has revolutionized our understanding of protoplanetary disks by revealing a large diversity of structures indicating concentrated dust growth, a better understanding and quantification of the dust growth processes can only be achieved with longer wavelength observations in the centimeter regime where the dust becomes optically thin, in particular through spectral index observations. Using the Square Kilometre Array (SKA) to derive dust spectral indices using even longer wavelengths ($>$5 cm) is impractical, as the dust emission becomes too faint to be detected at high spatial resolution. The ngVLA provides unprecedented possibilities to study dust growth in disks and further explore the dust trapping process.

%NOT SURE YET WHAT TO MAKE OF THIS RESULT IN THIS CONTEXT
%Interestingly, \citet{Pinilla2014} found that disk-integrated spectral indices of transition disks are systematically lower than those of primordial disks due to a lack of millimeter grains in the central part.

%To fill in this template, make sure that you read and follow the ASPCS Instructions for Authors and Editors available for download online.\footnote{Most URLs should be in a footnote like this one.  In this case, you can download the online material from \url{http://www.aspbooks.org}.}  Further hints and tips for including graphics, tables, citations, and other formatting helps are available there, in addition to the examples given in the aspauthor.tex file included here.  

%\acknowledgements ...  % Keep this text on the same line as the \verb"\acknowledgements" command because it makes things a lot easier.

%\bibliography{editor}  % For BibTex
%\bibliographystyle{apj}
%\bibliography{/Users/nienke/Dropbox/Research/myrefs.bib}

% For non-BibTex:

\end{document}